\documentclass[graybox]{svmult}

\usepackage{helvet}         
\usepackage{courier}        
\usepackage{type1cm}        
\usepackage{makeidx}         
\usepackage{graphicx}        
\usepackage{multicol}        
\usepackage[bottom]{footmisc}
\usepackage{amsmath,amssymb}

\newcommand{\FF}{\mathcal{F}}

\def\beq{\begin{equation}}
\def\eeq{\end{equation}}
\def\bea{\begin{eqnarray}}
\def\eea{\end{eqnarray}}
\def\ba{\begin{array}}
\def\ea{\end{array}}
\def\bi{\begin{itemize}}
\def\ei{\end{itemize}}

\def\bc{\begin{center}}
\def\ec{\end{center}}
\def\bs{\begin{small}}
\def\es{\end{small}}
\def\bfs{\begin{footnotesize}}
\def\efs{\end{footnotesize}}
\def\bt{\begin{tiny}}
\def\et{\end{tiny}}

\def\bd{\begin{document}}
\def\ed{\end{document}}
\def\lbl{\label}
\def\dst{\displaystyle}
\def\os{\overset}
\def\ub{\underbrace}

\def\vp{\vspace}
\def\hp{\hspace}

\begin{document}

\title*{On a nonlocal de Sitter gravity}
\author{Ivan Dimitrijevi\' c, Branko Dragovich, Zoran Raki\' c  and Jelena Stankovi\' c}
\institute{Ivan Dimitrijevi\' c \at Faculty of Mathematics, University of Belgrade, Studentski trg 16, p.o. box 550, 11000 Belgrade, Serbia \email{ivan.dimitrijevic@matf.bg.ac.rs}
\and Branko Dragovich \at Institute of Physics, University of Belgrade, Pregrevica 118, 11080 Belgrade, Serbia \email{dragovich@ipb.ac.rs}
\at Mathematical Institute of the Serbian Academy of Sciences and Arts, Kneza Mihaila 36, 11000 Belgrade, Serbia
\and Zoran Raki\' c \at Faculty of Mathematics, University of Belgrade, Studentski trg 16, p.o. box 550, 11000 Belgrade, Serbia \email{zoran.rakic@matf.bg.ac.rs}
\and Jelena Stankovi\' c \at Faculty of Education, University of Belgrade, Kraljice Natalije 43, Belgrade, Serbia \email{jelenagg@gmail.com}}
\maketitle

\abstract{
In this paper, we briefly review highlights  of nonlocal de Sitter gravity based on the nonlocal term  $ \sqrt{R - 2\Lambda}\ \mathcal{F}(\Box)\ \sqrt{R - 2\Lambda }$ in the Einstein-Hilbert action without matter sector.  This nonlocal de Sitter gravity model has several exact cosmological FLRW solutions  and one of these solutions contains some effects that are usually  assigned to dark matter and dark energy. There are also some other interesting and promising properties of this kind of gravity nonlocality. We also review some anisotropic cosmological solutions, and mention the corresponding nonlocal Schwarzschild-de Sitter metric. }

\section{Introduction}
 The Standard Model of Cosmology (SMC), or $\Lambda$CDM model, assumes that general relativity (GR) is  valid and applicable not only in the Solar system but also at the galactic and cosmological scale. Also there is almost common view that at the current cosmic time the universe approximately consists of 68 \% of  dark energy (DE), 27 \% of  dark matter (DM) and only  5 \% of well known   visible (ordinary) matter, see \cite{Planck2018}.
According to $\Lambda$CDM point of view, there exists a cold dark matter which is responsible for observational dynamics inside galaxies and their clusters, while dark  energy in the form of the cosmological constant $\Lambda$ acts as a repulsive force and causes accelerated expansion of the universe.  However, despite many efforts to confirm existence of DM and DE either in the sky or  in the laboratory experiments, they are still not discovered and remain hypothetical.

   In addition to problems with DE and DM, general relativity has its own problems, like singularity of the black holes solutions as well as singularity at  the beginning of the universe, and it means that GR should be modified in the vicinity of these singularities.  It is also well known that GR  is nonrenormalizable theory from the quantum point of view.  Although GR is one of the most beautiful and successful physical theories, which has given several significant  confirmed  predictions,
     it is nevertheless reasonable to doubt in its validity in description and understanding of all  astrophysical and cosmological gravitational phenomena.
Also, there is no a priory reason that GR should be appropriate gravity theory at all spacetime scales. Keeping all this in mind, it follows that general relativity is not a final theory of gravitation and that its extension is desirable.

 Since, there is still no  physical principle that could indicate in which direction we should search for right extension of GR, there are many approaches to its modification, see \cite{nojiri,clifton,nojiri1,capozziello,dragovich0} as some reviews. Despite many attempts, there is not yet generally accepted modification of general relativity. One of the current and attractive approaches is nonlocal modified gravity, see, e.g. \cite{biswas1,biswas3,biswas4,biswas5,deser,dimitrijevic13}. In a nonlocal  modification, the Einstein-Hilbert action is extended by a term that contains all higher order degrees of d'Alembert-Beltrami operator  $\Box = \nabla_\mu \nabla^\mu = \frac{1}{\sqrt{-g}} \partial_\mu (\sqrt{-g} g^{\mu\nu}\partial_\nu)$ usually  in the form of an analytic expression $F(\Box) = \sum_{n=0}^{+\infty} f_n \ \Box^n ,$ see \cite{koshelev2,koshelev3,eliz,koivisto,capozziello2009,dimitrijevic10}. There are also nonlocal gravity models  which contain some degrees of $\Box^{-1}$, e.g. see references \cite{maggiore,barvinsky2012,dimitrijevic13}.

One interesting class of nonlocal de Sitter models has the form
\begin{align} \label{eq1.1}
S = \frac{1}{16 \pi G} \int d^4 x \ \sqrt{-g}\ \big(R- 2 \Lambda  + P(R)\ \mathcal{F}(\Box)\ Q(R)\big) ,
\end{align}
where $\Lambda$ is the cosmological constant, $P(R)$ and $Q(R)$ are some differentiable functions of the scalar curvature $R$, and $\mathcal{F}(\Box)$ is an analytic function of $\Box$, see \cite{dimitrijevic1,dimitrijevic2,dimitrijevic3,dimitrijevic4,dimitrijevic5,dimitrijevic6,dimitrijevic7,dimitrijevic8,dimitrijevic9,dimitrijevic10,
dimitrijevic11,dimitrijevic12} and references therein. To better explore pure nonlocal effects, it is intentionally omitted term with matter in \eqref{eq1.1}.

This paper is a brief overview  of highlights of  nonlocal de Sitter gravity model \eqref{eq1.1}, where
\begin{align} \label{eq1.2}
P(R) = Q(R) = \sqrt{R - 2\Lambda} , \quad  \mathcal{F} (\Box) =  \sum_{n= 1}^{+\infty} \big( f_n \Box^n + f_{-n} \Box^{-n} \big) .
\end{align}
In particular, we will present several relevant exact vacuum cosmological solutions and some aspects of the corresponding Schwarzshild-de Sitter metric.
The first step in finding some exact cosmological solutions is  solving the equation $\Box \sqrt{R-2\Lambda} =  q \sqrt{R-2\Lambda} , $ where $ q =\eta \Lambda  \quad (\eta \in \mathbb{R})$ is an eigenvalue and  $\sqrt{R-2\Lambda}$ is an eigenfunction of the operator $\Box .$  One of these solutions mimics effects that are usually assigned to dark matter and dark energy. Some other solutions are examples of the nonsingular bounce ones in flat, closed and open universe. There are also singular and cyclic solutions. All these cosmological solutions are a result of nonlocality and do not exist in the local de Sitter case.

\section{Nonlocal de Sitter gravity: $\sqrt{dS}$ model}       
\label{Sec.2}

\subsection{Action}
 Our nonlocal de Sitter gravity model (introduced in \cite{dimitrijevic6}) is given by the action

\begin{align} \label{eq2.1}
S = \frac{1}{16 \pi G} \int d^4 x \ \sqrt{-g}\ \sqrt{R - 2\Lambda}\ F(\Box)\ \sqrt{R - 2\Lambda} ,
\end{align}
where $F (\Box)$ is the following formal expansion in terms of the d'Alemberian  $\Box$:
\begin{align} \label{eq2.2a}
 F(\Box) = 1+ \FF (\Box) = 1 + \FF_{+} (\Box) + \FF_{-} (\Box) ,  \\
 \FF_{+} (\Box) =\sum_{n=1}^{+\infty} f_n \ \Box^n , \ \FF_{-} (\Box) =\sum_{n= 1}^{+\infty} f_{-n} \ \Box^{-n}  \label{eq2.2b} .
 \end{align}
 When $F(\Box)= 1$, i.e. $\FF (\Box) = 0$, then  model \eqref{eq2.1} becomes
local de Sitter one and coincides with Einstein-Hilbert action with cosmological constant $\Lambda$:
\begin{align}\label{eq2.3a}
S_0 = \frac{1}{16 \pi G} \int d^4 x  \sqrt{-g}\ \sqrt{R - 2\Lambda} \sqrt{R - 2\Lambda} \\
= \frac{1}{16 \pi G} \int d^4 x \sqrt{-g}\ (R - 2 \Lambda) . \label{eq2.3b}
\end{align}
 It is worth pointing out that action \eqref{eq2.1} can be obtained in a very simple and natural way from action \eqref{eq2.3a}
 by embedding nonlocal operator \eqref{eq2.2a} within product $\sqrt{R - 2 \Lambda}\ \sqrt{R - 2 \Lambda} .$ By this  way,  $R$ and $\Lambda$ enter with the same form into nonlocal version as they are in the local one, and nonlocal operator $F(\Box)$ is dimensionless. In order to differentiate \eqref{eq2.1} from other non-local de Sitter models, we will often denote  model \eqref{eq2.1} as $\sqrt{dS} .$

\subsection{Equations of Motion}

 The  equations of motion (EoM) for model \eqref{eq1.1}, when $Q(R) =P(R)$, are given by (for more details, see \cite{dimitrijevic9,dimitrijevic13}):
\begin{align}
&G_{\mu\nu}+ \Lambda g_{\mu\nu} - \frac{g_{\mu\nu}}{2} P(R) \FF (\Box) P(R) + R_{\mu\nu} W - K_{\mu\nu} W + \frac 12 \Omega_{\mu\nu} = 0 , \label{eq2.7a}  \\
&W = 2 P'(R)\ \FF (\Box)\ P(R), \quad K_{\mu\nu} = \nabla_\mu \nabla_\nu - g_{\mu\nu}\Box ,  \label{eq2.7b}  \\
&\Omega_{\mu\nu} =  \sum_{n=1}^{+\infty} f_n \sum_{\ell=0}^{n-1} S_{\mu\nu}(\Box^\ell P, \Box^{n-1-\ell} P)
 -\sum_{n=1}^{+\infty} f_{-n} \sum_{\ell=0}^{n-1} S_{\mu\nu}(\Box^{-\ell-1} P, \Box^{-n+\ell} P), \label{eq2.7c}
\end{align}
where $S_{\mu\nu} (A, B)$ is defined as
\begin{align}
&S_{\mu\nu} (A, B) = g_{\mu\nu} \big(\nabla^\alpha A \ \nabla_\alpha B + A \Box B \big) - 2\nabla_\mu A\ \nabla_\nu B . \label{eq2.7d}
\end{align}

 If $P(R)$ is an eigenfunction of the corresponding d'Alembert-Beltrami operator $\Box$, and consequently also of its inverse $\Box^{-1}$, i.e.  holds, for $ q\neq 0 ,$
\bea \label{eq2.8}
\Box P(R) = q  P (R), \;   \Box^{-1} P(R) = q^{-1} P(R) ,   \; \FF (\Box) P(R) = \FF (q) P(R) ,
\eea
 then
\bea
  && \hp{-7mm} W = 2 \FF(q) P' P,  \quad \FF(q) = \sum_{n=1}^{+\infty} f_n \ q^n + \sum_{n=1}^{+\infty} f_{-n} \ q^{-n} , \label{eq2.8} \\[3pt]
 && \hp{-7mm} S_{\mu\nu}(\Box^{\ell} P, \Box^{n- 1 -\ell} P)  = q^{n-1} S_{\mu\nu} (P, P),  \label{eq2.8a} \\[7pt]
 && \hp{-7mm} S_{\mu\nu}(\Box^{-\ell-1} P, \Box^{-n+\ell} P) = q^{-n-1} S_{\mu\nu} (P, P),  \label{eq2.8b} \\[7pt]
 && \hp{-7mm} S_{\mu\nu}(P, P) = g_{\mu\nu} \big(\nabla^\alpha P \ \nabla_\alpha P + P \Box P \big) - 2\nabla_\mu P\ \nabla_\nu P,  \label{eq2.8b}\\[7pt]
 && \hp{-7mm}  \Omega_{\mu\nu} = \FF'(q) S_{\mu\nu}(P,P), \quad \FF'(q) = \sum_{n=1}^{+\infty} n\ f_n \ q^{n-1} -
 \sum_{n=1}^{+\infty} n\ f_{-n} \ q^{-n-1} \label{eq2.9}
\eea
and we get
\bea \label{eq2.10} \hp{-17mm}\ba{l}
  \dst{G_{\mu\nu}+ \Lambda g_{\mu\nu} - \frac{g_{\mu\nu}}{2}  \FF (q)P^2 + 2 \FF(q) R_{\mu\nu} P P' - 2 \FF(q) K_{\mu\nu} P P'}\\[10pt]\hp{17mm} \dst{ + \frac 12 \FF'(q) S_{\mu\nu}(P,P) = 0.}\ea
\eea
The last equation can be rewritten to
\bea  \hp{-10mm}\ba{l}\dst{
  \left(G_{\mu\nu}+ \Lambda g_{\mu\nu}\right)\left(1 + 2 \FF(q) P P'\right) + \FF(q)g_{\mu\nu}\left(-\frac 12 P^2 + P P'(R-2\Lambda)\right)}\\[7pt] \hp{27mm}\dst{- 2 \FF(q) K_{\mu\nu} P P' + \frac 12 \FF'(q) S_{\mu\nu}(P,P) = 0.}\ea \label{eq2.11}
\eea

 If $P(R)= \sqrt{R-2\Lambda}$, then  $P(R)P'(R) = \frac 12 $ and

\begin{equation}
\Box \sqrt{R - 2\Lambda} = q\,  \sqrt{R - 2\Lambda} = \eta\, \Lambda  \sqrt{R - 2\Lambda} , \qquad \eta\,\Lambda \neq 0 , \label{eq2.11a}
\end{equation}
where $q = \eta\, \Lambda$ and $q^{-1}= \eta^{-1} \Lambda^{-1}$ \ ($\eta$ -- dimensionless) follows from dimensionality of equalities \eqref{eq2.11a}.
Since $P(R)= \sqrt{R-2\Lambda}$,  EoM \eqref{eq2.11} simplify to
\bea  \label{eq2.12}
  \left(G_{\mu\nu}+ \Lambda g_{\mu\nu}\right)\left(1 + \FF(q)\right) + \frac 12 \FF'(q) S_{\mu\nu}(\sqrt{R-2\Lambda},\sqrt{R-2\Lambda}) = 0.
\eea

 It is clear that EoM \eqref{eq2.12} are satisfied if
\bea\label{eq2.13}
\mathcal{F} (q) = -1  \qquad \qquad\text{and} \qquad \qquad  \mathcal{F}' (q) = 0 .
\eea

It is worth pointing out that not only nonlocal de Sitter  model \eqref{eq2.1} is very simple and natural but also such are corresponding EoM \eqref{eq2.12}  with respect to all other models and their EoM that can be derived from  \eqref{eq1.1} with $\Lambda \neq 0$.

 Let us remark that nonlocal operator $\mathcal{F} (\Box)$, which satisfies conditions \eqref{eq2.13} in model \eqref{eq2.1},
can be taken in the rather general form \cite{dimitrijevic13}
\begin{align} \label{eq2.14}
 \mathcal{F}(\Box) =  \alpha e \ \frac{\Box}{q} \exp{\Big(-\frac{\Box}{q}\Big)} + \beta e\ \frac{q}{\Box} \exp{\Big(-\frac{q}{\Box}\Big)} , \quad \alpha + \beta = -1 ,    \quad q = \eta \Lambda .
\end{align}

\section{Cosmological Solutions}

\subsection{Cosmological Solutions in Homogenous and Isotropic Space}

At the cosmological scale the universe is homogeneous and isotropic with
the Friedmann-Lema$\mathrm{\hat{i}}$tre-Robertson-Walker (FLRW) metric, $(c=1 , \,  k= 0, \pm 1)$,
\begin{align}
ds^2 = - dt^2 + a^2(t)\left(\frac{dr^2}{1-k r^2} + r^2 d\theta^2 + r^2 \sin^2 \theta d\phi^2\right)  . \label{eq3.1}
\end{align}

For the FLRW metric \eqref{eq3.1} we have the following expressions for scalar curvature $R$ and operator $\Box$:
\begin{align}
R (t) &= 6\Big(\frac{\ddot a}{a} + \big(\frac{\dot a}{a}\big)^2 +\frac{k}{a^2}\Big),  \label{eq3.2} \\
\Box R &= - \frac{\partial^2}{\partial t^2} {R}-3 H \frac{\partial}{\partial t} {R},   \label{eq3.3}
\end{align}
where $H= \dst{\frac{1}{a} \frac{d a}{d t}\,  \equiv \frac{\dot{a}}{a}}$ is the Hubble parameter.


In the sequel of this subsection, we present some exact cosmological solutions with $\Lambda \neq 0 .$ For some details, see
\cite{dimitrijevic6,dimitrijevic10,dimitrijevic13} and references therein.

\subsubsection{Solutions  of the form $a(t) = A\ t^n\ e^{\gamma t^2}, \  (k= 0)$ }

In the case of next eight flat spaces (k=0), see \cite{dimitrijevic13} .

There are two solutions of the form
\begin{align}
a(t) = A\ t^n\ e^{\gamma t^2} ,  \quad k= 0 ,  \label{eq3.4}
\end{align}
where $ n$ and $\gamma$ are some definite real constants.

The eigenvalue problem
\begin{equation}
  \Box \sqrt{R-2\Lambda} = q \sqrt{R-2\Lambda} , \qquad   q = \eta \Lambda \neq 0   \label{eq3.5}
\end{equation}
is satisfied in the following two cases:
\begin{align}
1.& \quad n =\frac{2}{3} , \qquad \gamma = \frac{\Lambda}{14} , \qquad   q = - \frac{3}{7}\Lambda , \label{eq3.6a} \\
2.& \quad   n=0 , \qquad \gamma = \frac{1}{6}\Lambda , \qquad  q = -\Lambda .  \label{eq3.6b}
\end{align}

Using \eqref{eq3.6a} and \eqref{eq3.6b}, we obtain the following two solutions in flat space:
\begin{align}
 &a_1(t)  = A\ t^{\frac{2}{3}}\ e^{\frac{\Lambda}{14} t^2} , \quad k =0, \quad \mathcal{F}(-\frac{3}{7} \Lambda) = -1 , \ \ \mathcal{F}'(-\frac{3}{7} \Lambda) = 0 ,  \label{eq3.7} \\
 &a_2(t)  = A\  e^{\frac{\Lambda}{6} t^2}, \quad k =0, \quad \mathcal{F}(-\Lambda) = -1 , \ \ \mathcal{F}'(- \Lambda) = 0 .  \label{eq3.8}
\end{align}

\subsubsection{Solutions  of the form $a(t) = (\alpha \ e^{\lambda t} + \beta \ e^{-\lambda t})^\gamma$, \ ($k=0$)}
\label{3.2.2}

We have the following  two special solutions:
\begin{align}
 &a_3 (t) = A\ \cosh^{\frac{2}{3}}{\big(\sqrt{\frac{3}{8} \Lambda}\ t\big)} , \quad  k= 0 , \quad \FF\big(\frac{3}{8} \Lambda\big)=-1, \ \FF'\big(\frac{3}{8} \Lambda\big)=0 , \label{eq3.9} \\
 &a_4 (t) = A\ \sinh^{\frac{2}{3}}{\big(\sqrt{\frac{3}{8} \Lambda}\ t\big)} , \quad  k= 0 ,  \quad \FF\big(\frac{3}{8} \Lambda\big)=-1, \ \FF'\big(\frac{3}{8} \Lambda\big)=0 . \label{eq3.10}
\end{align}

\subsubsection{Solutions  of the form $a(t) = (\alpha \ \sin{\lambda t} + \beta \ \cos{\lambda t} )^\gamma$, \ ($k=0$)}

In this case, we obtained the following four solutions:
\begin{align}
 &a_5 (t) = A\ \Big(1 + \sin{\big(\sqrt{-\frac{3}{2} \Lambda}\ t\big)} \Big)^{\frac{1}{3}}, \
           k= 0 , \quad \FF\big(\frac{3}{8} \Lambda\big)=-1, \ \FF'\big(\frac{3}{8} \Lambda\big)=0 ,  \label{eq3.11} \\
 &a_6 (t) = A\ \Big(1 - \sin{\big(\sqrt{-\frac{3}{2} \Lambda}\ t\big)} \Big)^{\frac{1}{3}}, \
           k= 0 , \quad \FF\big(\frac{3}{8} \Lambda\big)=-1, \ \FF'\big(\frac{3}{8} \Lambda\big)=0 , \label{eq3.12} \\
 &a_{7} (t) = A\  \sin^{\frac{2}{3}}{\big(\sqrt{-\frac{3}{8} \Lambda}\ t\big)} , \quad   k= 0 , \quad \FF\big(\frac{3}{8} \Lambda\big)=-1, \ \FF'\big(\frac{3}{8} \Lambda\big)=0 ,  \label{eq3.13} \\
 &a_{8} (t) = A\  \cos^{\frac{2}{3}}{\big(\sqrt{-\frac{3}{8} \Lambda}\ t\big)}   ,
      \quad     k= 0 , \quad \FF\big(\frac{3}{8} \Lambda\big)=-1, \ \FF'\big(\frac{3}{8} \Lambda\big)=0 . \label{eq3.14}
\end{align}

\bigskip

We have three type of vacuum solutions in the closed and open FLRW space presented in the next two subsubsections.

\subsubsection{Cosmological solution  of the form $a(t) = A\ e^{\pm \sqrt{\frac{\Lambda}{6}} t}$,  ($k = \pm 1$)}

In the paper \cite{dimitrijevic10}, we presented the following exact solution:
\begin{align} \label{eq3.15}
a_{9}(t) = A\ e^{\pm \sqrt{\frac{\Lambda}{6}} t} , \quad k = \pm 1, \quad \mathcal{F}(\frac{1}{3} \Lambda) = -1 , \ \ \mathcal{F}'(\frac{1}{3} \Lambda) = 0,
\quad \Lambda > 0.
\end{align}
Note that this solution is different with respect to the de Sitter one.

\subsubsection{Solutions of the form $a(t) = (\alpha \ e^{\lambda t} + \beta \ e^{-\lambda t})^\gamma$, \ ($k=\pm 1$)}

When $\alpha \neq 0, \  \beta \neq 0 , \ R\neq 2\Lambda, \ q \neq 0$ and $k \neq 0$  we obtain
\begin{equation}
 \gamma=\frac{1}{2},\qquad  q=\frac{1}{3} \Lambda, \qquad  \lambda= \pm \sqrt{\frac{2}{3} \Lambda} ,\qquad  k \neq 0 . \label{eq3.16}
\end{equation}
The corresponding cosmological solutions are:
\begin{align}
&a_{10} (t) = A\ \cosh^{\frac{1}{2}}{\big(\sqrt{\frac{2}{3} \Lambda}\ t\big)} , \quad  k= \pm 1 ,  \quad \FF\big(\frac{1}{3} \Lambda\big)=-1, \ \FF'\big(\frac{1}{3} \Lambda\big)=0 , \label{eq3.17} \\
 &a_{11} (t) = A\ \sinh^{\frac{1}{2}}{\big(\sqrt{\frac{2}{3} \Lambda}\ t\big)} , \quad  k= \pm 1 , \quad \FF\big(\frac{1}{3} \Lambda\big)=-1, \ \FF'\big(\frac{1}{3} \Lambda\big)=0 .  \label{eq3.18}
\end{align}

\subsection{Cosmological Solutions in Anisotropic Space}

We are going now to present some solutions for the Bianchi type I anisotropic metric in the form (for details, we refer to \cite{dimitrijevic14})
\begin{equation}  \label{eq3.19}
  \mathrm{d}s^2 = -\mathrm d t^2 +a(t)^2\left(e^{2\beta_1(t)}\mathrm dx^2+ e^{2\beta_2(t)}\mathrm dy^2 + e^{2\beta_3(t)}\mathrm dz^2\right)
\end{equation}
with condition
\begin{equation}
  \beta_1(t)+ \beta_2(t) + \beta_3(t) =0.   \label{eq3.20}
\end{equation}

Let us introduce  $\sigma(t)$ as follows:
\begin{equation}
  \sigma(t)^2 = \dot \beta_1(t)^2+ \dot \beta_2(t)^2 + \dot \beta_3(t)^2.   \label{eq3.21}
\end{equation}

One can obtain the following expressions
\begin{align}
  R &= R_{FLRW} + \sigma^2, \label{eq3.22a}\\
  \Box u(t)&= \Box_{FLRW}u(t), \label{eq3.22b}
\end{align}
where index $FLRW$ denotes quantities corresponding to the $FLRW$ metric with scale factor $a(t)$ and $k=0$.\\

Solving the eigenvalue problem
\begin{align} \label{eq3.23}
\Box\ \sqrt{R - 2 \Lambda} = q \ \sqrt{R - 2 \Lambda} ,
\end{align}
we obtain several solutions for scale factor $a(t)$ together with  function $\sigma^2 (t)$, with the  corresponding conditions $\mathcal{F} (q) = -1$ and $\mathcal{F}' (q) = 0$. For flat space $(k=0)$ and constant  $\sigma^2$ we have:
\begin{align}
&a_1(t) = A\ t^{2/3}e^{\frac\Lambda{14}(1-\eta) t^2} ,  \quad q = -\frac{3}{7} \Lambda (1-\eta), \quad  \sigma^2 = 2\Lambda \eta  , \label{eq3.24a} \\
&a_2(t) = A \cosh^{\frac{2}{3}}(\sqrt{\frac{3 \Lambda}{8}} (1-\eta) \; t)  , \quad
  q = \frac{3 \Lambda}{8} (1-\eta)^{2}, \quad \sigma^{2} = 2 \Lambda \eta (2-\eta),  \label{eq3.24b} \\
&a_3(t) = A \sinh^{\frac{2}{3}}(\sqrt{\frac{3 \Lambda}{8}} (1-\eta) \; t)  , \quad
  q = \frac{3 \Lambda}{8} (1-\eta)^{2}, \quad \sigma^{2} = 2 \Lambda \eta (2-\eta), \label{eq3.24c}
  \end{align}
\begin{align}
&a_4(t) = A \cos^{\frac{2}{3}}(\sqrt{-\frac{3 \Lambda}{8}} (1-\eta) \; t)  , \quad
  q = \frac{3 \Lambda}{8} (1-\eta)^{2}, \quad  \sigma^{2} = 2 \Lambda \eta (2-\eta),  \label{eq3.24d} \\
&a_5(t) = A \sin^{\frac{2}{3}}(\sqrt{-\frac{3 \Lambda}{8}} (1-\eta) \; t)  , \quad
  q = \frac{3 \Lambda}{8} (1-\eta)^{2}, \quad  \sigma^{2} = 2 \Lambda \eta (2-\eta).  \label{eq3.24e}
\end{align}

Note that the above anisotropic solutions \eqref{eq3.24a} -- \eqref{eq3.24e} tend to the corresponding homogeneous and isotropic ones presented in the previous subsection, when $\sigma^2 \to 0\ (\eta \to 0).$

\section{Concluding Remarks}
\label{sec:4}

From the Friedmann equations, it is useful to introduce effective energy density $\bar{\rho}$ and  pressure $\bar{p}$:
\begin{equation}
\bar{\rho}(t) = \frac{3}{8\pi G} \Big(\frac{\dot{a}^2 + k}{a^2} - \frac{\Lambda}{3}\Big) , \quad \bar{p} (t) = \frac{1}{8\pi G}
\Big(\Lambda - 2\frac{\ddot{a}}{a} - \frac{\dot{a}^2 + k}{a^2}\Big) , \label{eq4.1}
\end{equation}
and the  equation of  state
\begin{equation}
\bar{p} (t) = \bar{w}(t) \, \bar{\rho} (t) , \label{eq4.2}
\end{equation}
where $\bar{w}(t)$ is the corresponding  effective state parameter. Using \eqref{eq4.1} and \eqref{eq4.2}, one can compute $\bar{\rho}(t)$ and
$\bar{p} (t)$ for each of the above cosmological solutions $a(t)$. In this way, it is shown in \cite{dimitrijevic10,dimitrijevic13} that solution
$a(t) = A\ t^{\frac{2}{3}}\ e^{\frac{\Lambda}{14} t^2}$ mimics dark energy and dark matter.

Recently, we started to investigate Schwarzschild-de Sitter spacetime in $\sqrt{dS}$ nonlocal gravity with metric
\begin{align} \label{eq4.3}
ds^2 = -A(r)\ dt^2 + A^{-1}(r)\ dr^2 + r^2\ d\theta^2 + r^2\ \sin^2{\theta}\ d\varphi^2\ .
\end{align}
The corresponding eigenvalue problem
\begin{align} \label{eq4.4}
\Box \sqrt{R(r)- 2\Lambda}  = \frac{1}{r^2} \frac{\partial}{\partial r}\big[r^2 A(r) \frac{\partial }{\partial r} \sqrt{R(r)- 2\Lambda}  \big] = q\, \sqrt{R (r)- 2\Lambda}
\end{align}
is rather nonlinear in $A(r)$, since unknown function $A(r)$ is contained also in the scalar curvature $R(r)$, i.e $R(r) =\frac{1}{r^2} \frac{\partial^2}{\partial r^2}\big[r^2 \big(1-A(r) \big)\big]$. We found an approximative  solution of \eqref{eq4.4}, while finding exact solution remains a challenge, see \cite{dimitrijevic15}.

It is worth noting that action \eqref{eq2.1} of nonlocal $\sqrt{dS}$ gravity can be transformed to
\begin{align}\label{eq4.5}
S = \frac{1}{16\pi G} \int [R- 2\Lambda + (R - 4\Lambda)\ \mathcal{F}(\Box)\ (R- 4\Lambda)] \ \sqrt{-g} \ d^4x
\end{align}
when $|R| \ll |2 \Lambda|$, for details, see \cite{dimitrijevic12}. Nonlocal gravity model \eqref{eq4.5} contains cosmological solution $a(t) = A \sqrt{t} e^{\frac{\Lambda}{4} t^2}$ \cite{dimitrijevic11} which mimics an interplay between radiation $\sqrt{t}$ and dark energy $e^{\frac{\Lambda}{4} t^2}$. It also has some vacuum  solutions that change topology when local de Sitter gravity extends by this nonlocal one \cite{dimitrijevic12}.

It is also worth noting that employment of nonlocal operator in the form of exponential function comes from string theory, in particular $p$-adic string theory \cite{dragovich1}. Nonlocality in the matter sector is also considered, see, e.g. \cite{dragovich2,dragovich3,arefeva2,koshelev2011}.

Ending, we can conclude that so far obtained results in $\sqrt{dS}$ nonlocal gravity are encouraging. We plan to explore  other aspects of $\sqrt{dS}$, in particular inflation, gravitational waves and   inclusion of
 matter sector.

\section*{Acknowledgments}

This research was partially funded by the Ministry of Education, Science and Technological Developments of the Republic of Serbia: grant number 451-03-47/2023-01/ 200104 with University of Belgrade, Faculty of Mathematics, and grant number 451-03-1/2023-01/4 with Faculty of Education, University of Belgrade. It is also partially supported by the COST Action: CA21136 – Addressing observational tensions in cosmology with
systematics and fundamental physics (CosmoVerse). I.D. is thankful to prof. V. Dobrev for hospitality during the conference LT- 15 in Varna.

\end{document}